\documentclass[aps,floatfix,twocolumn,groupedaddress,showpacs,groupauthors]{revtex4}
\usepackage{graphics}
\usepackage{epsfig}
\usepackage{graphicx}
\usepackage{dcolumn}

\begin{document}

\title{Large magnetoresistance in La$_{2/3}$Ca$_{1/3}$MnO$_{3}$ thin films induced by metal masked ion damage technique}

\author{M. J. Zhang, J. Li, Z. H. Peng, D. N. Zheng
\email{dzheng@ssc.iphy.ac.cn}} \affiliation{National laboratory for
Superconductivity, Beijing National Laboratory for Condensed Matter
Physics, and Institute of Physic, Chinese Academy of Sciences,
Beijing 100080, People's Republic of China}
\author{A. Z. Jin, C. Z. Gu}
\affiliation{Laboratory of Microfabrication, Beijing National
Laboratory for Condensed Matter Physics, and Institute of Physic,
Chinese Academy of Sciences, Beijing 100080, People's Republic of
China}

\begin{abstract}
{We have developed a simple process to obtain large
 magnetoresistance (MR) in perovskite manganite thin films by a
 combination of focused ion beam (FIB) milling and 120 keV H$_{2}^{+}$ ion implantation.
 Metal slits about 70 nm in width were printed by 30 kV focused Ga ion beam
 nanolithography on a 4 mm track, and the materials in these slits are then
 irradiated by the accelerated H$_{2}^{+}$ ions. Using this method, in a magnetic
 field of 5 T we can get a MR${>}$60\% over a 230 K temperature scope, with a
 maximum value of 95\% at around 70 K. This technique is very promising in
 terms of its simplicity and flexibility of fabrication and has potential
 for high-density integration.}
\end{abstract}
 \maketitle

\section{INTRODUCTION}
The so-called colossal magnetoresistance (CMR) has been extensively
studied recently. Very large resistance changes of about 10$^{6}\%$
have been obtained in applied magnetic fields of several Tesla for a
range of different perovskite-like manganites
\cite{Ramirez,Subramanian,Jin,Helmolt}. The MR phenomenon has been
divided into intrinsic and extrinsic: whereas intrinsic effects are
found in bulks of ferromagnetic materials and are determined by
material parameters, extrinsic effects are found only at defect
structures, suitable artificial heterostructures and devices. To
obtain a large MR in a relatively low magnetic field, many research
groups focus on the investigation of extrinsic MR in various
magnetic oxides in recent years. Extrinsic MR in ferromagnetic
oxides usually falls into three broad classes, namely grain-boundary
MR \cite{Gupta,Mathur}, tunnel junction MR \cite{Lu} and domain wall
MR \cite{Wu,Wolfman}. In this article we report a method aimed to
enhance the MR in the LCMO thin films by forming nano-constraints
through a combination of focused ion beam (FIB) milling and 120 KeV
H$_{2}^{+}$ ion implantation. The results demonstrate that a maximum
MR ratio of $95\%$ (defined as $\triangle
R/R_{0}=(R_{H}-R_{0})/R_{0}$, where $R_{0}$ is the resistance at a
magnetic field of H=0 and $R_{H}$ at H=5T.) and an overall value
$>65\%$ can be reached in a 230K temperature range.

\section{EXPERIMENT}
\begin{figure}
\includegraphics[width=1\columnwidth]{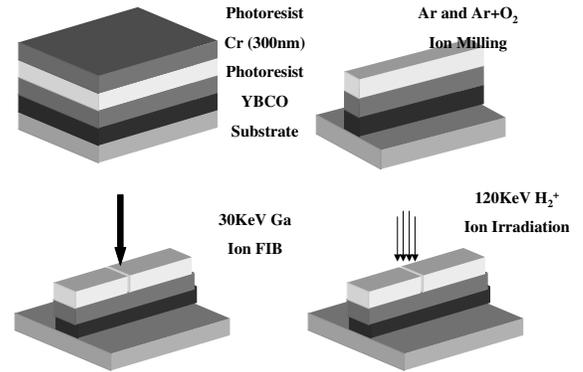}
\caption{\label{fig1}Schematic diagram of the fabrication process to
create highly localized barriers.}
\end{figure}

\begin{figure}
\includegraphics[width=.8\columnwidth]{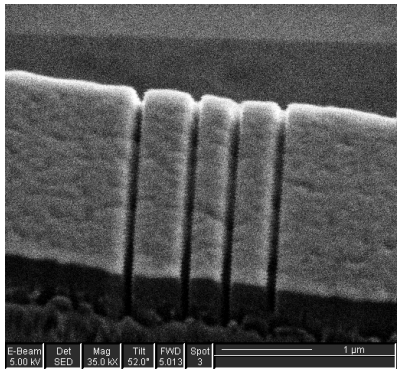}
\caption{\label{fig2}Secondary electron image obtained in the FIB of
4 slits approximately 70nm wide cutting into a typical metal mask.}
\end{figure}

The films for this study were high quality c-axis oriented 100 nm
thick epitaxial La$_{2/3}$Ca$_{1/3}$MnO$_{3}$ (LCMO) grown on (100)
LaAlO${_3}$ substrates by pulsed laser ablation. A tri-layer mask,
consisting of 900 nm S9918 photo resist, 300 nm metal Cr deposited
by dc magnetron sputtering and another 900 nm S9918 photo resist,
was then coated on the sample. As shown schematically in Fig.
\ref{fig1}a, tracks with a nominal width of 4 mm were patterned by
optical lithography and Ar ion milling at 400 eV and 10 mA on a
water-cooled rotating stage. When the figure was transferred from
photoresist to the Cr film, we adopted reactive ion etching (RIE) to
pattern the track of 4 mm, as shown in Fig. \ref{fig1}b. The Cr
masking layer was chosen to be sufficiently thick to absorb the
implanted protons during the following implantation procedure, but
also thin enough that the slit could be cut with sufficient
accuracy. To prepare the mask apertures, the patterned chip was
mounted on a carrier that was transferred to the FIB system with a
Ga source. Using 30 KeV Ga ions, slots of single scan-line width
were milled at 10 pA in the Cr film, as can be seen in Fig.
\ref{fig1}c. Earlier experimental results showed that Ga ions would
be implanted into the slit materials during the milling process,
which should be avoided \cite{Park}. For this reason, we chose S9918
photoresist as a barrier layer to absorb the Ga ion. In order to
make sure the material in the slits is now LCMO, we again adopted
RIE to remove the remaining photoresist in the slits. The last step,
H$_{2}^{+}$ irradiation, was then preceded to create highly
localized barriers. This process is shown in Fig. \ref{fig1}d. Fig.
\ref{fig2} shows a secondary electron image obtained in the FIB
system, with 4 slits approximately 70 nm wide cutting into a Cr
metal mask.

The resistivity and magnetoresistance of the as prepared junction
were measured by the four-probe method in the temperature range 5
K$<$T$<$300 K. An MPMS superconducting quantum interference (SQUID)
measuring system was used to generate the uniform magnetic field.

\section{RESULTS AND DISCUSSION}
\begin{figure}
\includegraphics[width=.9\columnwidth]{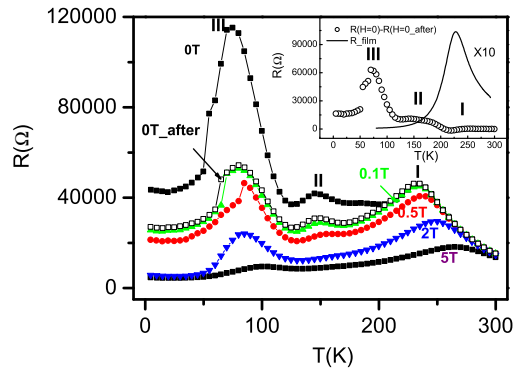}
\caption{\label{fig3}Resistance versus temperature curves for the
LCMO junction in zero field and fields of 5T, 2T, 0.5T, and 0.1T.
The curve measured with the field totally withdrawn is also plotted.
The inset shows the R-T curves of an epitaxial film and the
resistance change between the initial and the after-measurement
values of the junction.}
\end{figure}

The effect of our processing method on the transport properties in
the junction is evidenced in the temperature dependence of the
resistance (in Fig. \ref{fig3}). The original film (inset of Fig.
\ref{fig3}) exhibits a characteristic insulator-metal (I-M)
transition peak at around 230 K, which is also a signature of its
paramagnetic-ferromagnetic transition. The junction, however, shows
three peaks on its resistance versus temperature curve. The first
peak is consistent with the I-M transition in the LCMO thin film,
and is obviously contributed by the film portions other than the
irradiated slits. The H$_{2}^{+}$ implantation intuitionally reduces
the LCMO film, makes the slits oxygen deficient, and therefore
shifts the I-M transition point left to the second peak position.
The origin of the third peak is still not known. The data
fluctuation around this peak suggests that it may be owe to a kind
of metastable state. We are concerned that it is a result of spin
pinning at the local barriers, and concomitantly of large spin
angles.

When a 5T magnetic field was applied, the sample resistance was
remarkably decreased. The first and the third peak clearly right
shift to higher temperatures and the second one is completely
suppressed, owe to the reduction of spin fluctuations in a magnetic
field. With the field decreasing from 5T to 2T, 0.5T, and 0.1T, the
second peak gradually resumes. However, the low temperature
resistance never recovers to its original value even when the field
is totally withdrawn and the sample is warmed to room temperature to
remove any possible magnetic remanence. The resistance change
between the initial and the after-measurement values is also shown
in the inset, whose apex is coincident with the position of the
third resistance peak. This again suggests that the low temperature
peak is from a metastable state, which may break down in a high
field and thus be irreversible.

\begin{figure}
\includegraphics[width=.85\columnwidth]{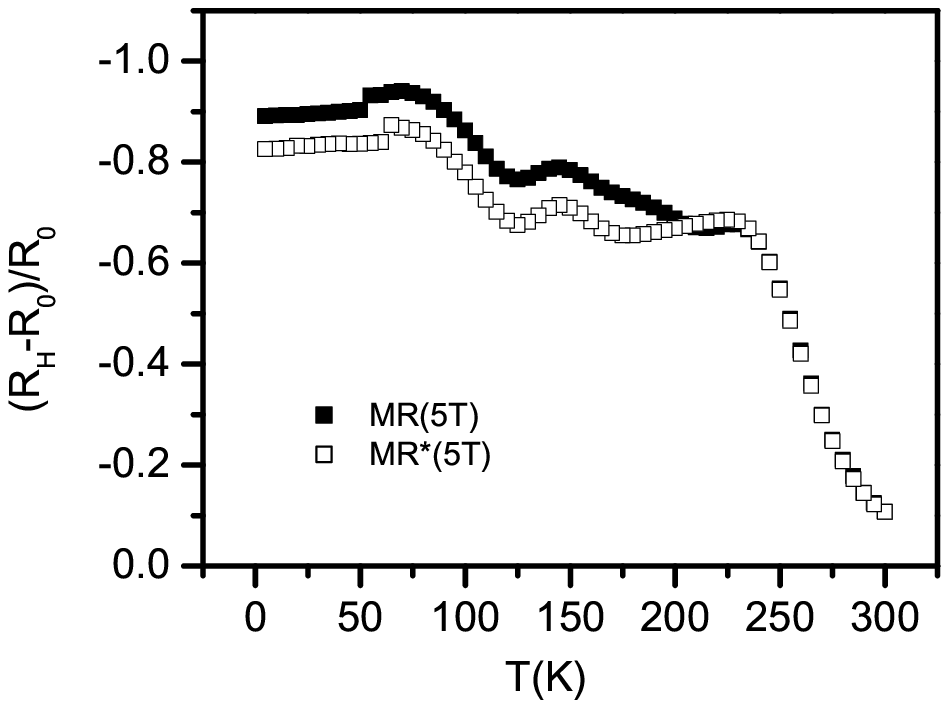}
\caption{\label{fig4}The MR ratio ($\triangle R/R_{0}$) and the
MR$^{*}$ ratio ($\triangle R/R_{0}$-after) versus temperature
curves.}
\end{figure}

The temperature dependence of the MR ratio (defined as $\triangle
R/R_{0}=(R_{H}-R_{0})/R_{0}$) in 5T field is read from Fig.
\ref{fig3} and plotted in Fig. \ref{fig4}. It is impressive that the
MR ratio is above 60\% in temperatures below 230K. Three small bumps
appear at temperatures corresponding to the resistance peaks. A
maximum value as large as 95\% is obtained at 70K, near the third
resistance peak. Another ratio denoted as MR$^{*}$ is also plotted
in the figure, where the after-measurement resistance is adopted
instead of R$_{0}$. Apparently, MR$^{*}$ evolves like MR, besides
the slightly reduced value as low temperatures.

\begin{figure}
\includegraphics[width=.9\columnwidth]{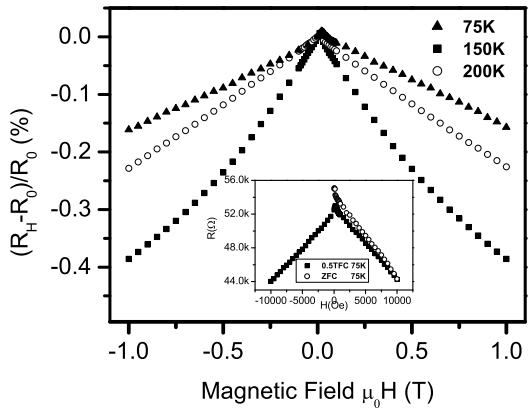}
\caption{\label{fig5}MR ratios $\triangle R/R_{0}$ of the sample at
75K, 150K and 200K in a -1T to 1T field regime. The inset shows the
difference between zero-field-cool and 5000Oe-field-cool MRs at 75K}
\end{figure}

To probe the nature of the greatly enhanced MR in the junction, we
recorded the sample resistance in a magnetic field scanning from -1T
to +1T at temperatures 200, 150, and 75K. The results are shown in
Fig. \ref{fig5}. It is noticed that, at the three temperatures, the
junction resistance all shows an almost linear dependence on the
magnetic field, with no low field shoulders and ignorable
hystereses, suggesting an intrinsic behavior. The MR ratio in 1T
field is 24\% and 39\% at 200K and 150K, respectively. The MR
enhancement at 150K is most probably due to the irradiation induced
spin disordering at the slits. The MR ratio at 75K is only 17\% in
1T field, which is somewhat under-estimated because the zero field
resistance cannot recover after measured in a 5T magnetic field.
Nevertheless, the value is still substantial as compared with
epitaxial thin films. In the case of zero-field-cool, the MR ratio
increases to about 20\%.

Therefore, the transport measurement results reveal that the device
fabricated does not form nanoconstraints or magnetic domain walls in
the ion irradiated slits as we had expected. In stead, the
characters of the material in the damaged region are changed upon
the ion implantation. The width of the slit (~70 nm) might be too
large to become a geometric constraint for domain walls, or the
tunneling barriers. Nevertheless, the large MR ratio obtained in
such a broad temperature scope still makes the device potential for
eventual technological applications.

\section{CONCLUSIONS}

In conclusion, we have developed a metal-masked ion-damage method to
improve the MR effect in LCMO thin films. A large MR ratio over 60\%
was observed in 5T field in a 230 K temperature scope, with a
maximum value of 95\% at 70K. The linear field dependence of the MR
ratio reveals its intrinsic nature. Although the origin of the large
MR ratio at low temperatures still has not been clarified, the
device demonstrates a very promising technique for practical
applications, in terms of its simplicity and flexibility of
fabrication and its potential for high-density integration.

\section{ACKNOWLEDGMENTS}

The authors would like to thank Mr. W. W. Huang and Mrs S. L. Jia
for the transport measurements.This project is sponsored by the
National Science Foundation of China under grant Nos.10574154,
50472076, and 10221002, Ministry of Science and Technology of China
(2006CB601007), Chinese Academy of Sciences, and National Center for
Nanoscience and Technology, China.

\end{document}